\documentclass{ckm}
\usepackage{mathptmx} 

\usepackage{epsf}
\newcommand{\be}{\begin{equation}}
\newcommand{\ee}{\end{equation}}
\newcommand{\bea}{\begin{eqnarray}}
\newcommand{\eea}{\end{eqnarray}}

\confname{Workshop on the CKM Unitarity Triangle, IPPP Durham, April
  2003}

\title{ CP asymmetry in $B^- \to \pi^+ \pi^- K^-$ 
   and  $B^- \to K^+ K^- K^-$    decays }

\author{T. N. Pham\addressmark{a}}

\address[a]{ Centre de Physique Theorique, \\
Centre National de la Recherche Scientifique, UMR 7644,  \\  
Ecole Polytechnique, 91128 Palaiseau Cedex, France}

\begin{document}

\begin{abstract}
The near equality between the measured value for $|V_{ub}|/|V_{cb}|$ and
its unitarity triangle lower bound from $\sin 2\beta$  measurements suggests 
that $\alpha \approx 90^{\circ}$
and $\sin\gamma \approx \cos\beta = 0.91$ and a large direct CP
violation in $B$ decays would be possible. 
In this talk, I would like to discuss 
a recent analysis of the non resonant and resonant CP asymmetry 
for the decays  $B^- \to \pi^+ \pi^- K^-$
and $B^- \to K^+ K^- K^-$ where  the CP asymmetry at the $\chi_{c0}$
resonance comes from the interference of the
non resonant amplitude with the resonant amplitude
$B^\pm \to \chi_{c0} K^\pm $ $ \to\pi^+ \pi^- K^\pm $ and 
$B^\pm \to \chi_{c0} K^\pm$ $ \to K^+ K^- K^\pm$. Using the
 the Babar and Belle measurements of the 
$B^-\to \chi_{c0} K^- $ branching ratios, we predict that 
partial width asymmetry at the   $\chi_{c0}$ resonance
could reach  $10 \%$ for 
the $B^- \to \pi^+ \pi^- K^-$ decay
and $16\%$ for the  $B^- \to K^+ K^- K^-$decay.
\end{abstract}

\maketitle

\section{Introduction}

In three-body charmless $B$ decays, e.g $B\to K\pi\pi $, apart from the
non resoant CP asymmetry there is also the partial width
CP asymmetries produced by the interference between the 
non resonant and resonant amplitudes, both of them could be tree amplitudes.
Previous studies \cite{Eilam,Deshpande,Fajfer} show that in 
$B^- \to \pi^+ \pi^- \pi^-$ decays, a large  partial width CP asymmetry  
at the  $\chi_{c0}$ resonance  could 
arise from the interference of the non resonant amplitude 
with the resonant amplitude coming from the decays 
$B^{-}\to \chi_{c0}\pi^{-}$ $\to$ $\chi_{c0}\to \pi^- \pi^-$ , but  no 
firm prediction for the asymmetry could be given as there is no theoretical
prediction for the $B^{-}\to \chi_{c0}\pi^{-}$ decay rate. A similar 
analysis can now be made for the CP asymmetries in the 
$B^-\to \pi^+ \pi^- K^- $ and $B^-\to K^+ K^- K^- $ decays using the recent 
Belle\cite{Belle1} and  Babar\cite{Babar} measurements of the $B^-\to \chi_{c0} K^- $
branching ratios:  
${\rm BR}(B^-\to \chi_{c0} K^-) = (6.0^{+2.1}_{-1.8})\times 10^{-4}$
${\rm BR}(B^-\to \chi_{c0} K^-) = (2.4\pm 0.7)\times 10^{-4}$ 
for Belle and Babar respectively. The non resonant branching 
ratios reported by Belle\cite{Belle2} are:
${\rm BR}(B^-\to \pi^+ \pi^- K^-) =(58.5\pm 7.1 \pm 8.8) \times
10^{-6}$
${\rm BR}(B^-\to K^+ K^- K^-) = (37.0\pm 3.9 \pm 4.4) \times 10^{-6}$.
In this talk, I would like to discuss a recent work\cite{Fajfer1}
on the non resonant CP asymmetries and the partial CP asymetries 
at the $\chi_{c0}$ reosnance in these decays.

\section{The CKM matrix}
The effective weak Hamiltonian  for the nonleptonic Cabibbo-suppressed
$B$ meson decays is given by\cite{Ali,Deshpande1,Isola}
\begin{eqnarray}
&&\kern -0.7cm  {\mathcal H}_{\rm eff}\kern -0.1cm =\kern -0.1cm  \frac{G_F}{{\sqrt 2}} 
[V_{ub} V_{us}^*(c_1 O_{1u} \kern -0.1cm+\kern -0.1cm c_2 O_{2u} ) 
 +  V_{cb} V_{cs}^* (c_1 O_{1c} +c_2 O_{2c} ) ]\nonumber\\
&&\kern -0.7cm - \sum_{i=3}^{10} ([V_{ub} V_{us}^* c_i^u
     +  V_{cb} V_{cs}^* c_i^c + V_{tb} V_{ts}^* c_i^t) O_i ] + \rm h.c.
\label{eq1}
\end{eqnarray}
where $O_{1}, O_{2}$ are the tree-level operators and $O_{3}-O_{6}$
are the penguin operators. As usual with the factorization model we use
in our analysis, the hadronic marix elements are obtained from the 
effective Hamiltonian in  
Eq.(\ref{eq1}) with $c_{i}$ replaced by $a_{i}$ where $c_{i}$
are next-to-leading Wilson coefficients. Since  $a_3$ and $a_5$
are one order of magnitude smaller than $a_4$ and $a_6$, the contributions
from $O_{3}$ and $O_{5}$ can be safely neglected. For
$N_{c}=3$, $m_b = 5 \rm \, GeV$, we use \cite{Deshpande1,Isola}~: 
$a_1= 1.05$ ,  $a_2=  0.07$ , $ a_4=-0.043- 0.016\,i$, 
$a_6=  -0.054- 0.016\,i$. Before writing down the decay amplitudes, I would
like to discuss the current determination of the CKM parameters 
and show that the unitarity triangle lower bound for 
$|V_{ub}|/|V_{cb}|$ obtained
from the Babar and Belle measurements of $\sin 2\beta$ and the measured
value for $|V_{ub}|/|V_{cb}|$ already tells us that 
$\alpha \approx 90^{\circ}$ and $\sin \gamma \approx \cos\beta =0.91$ .
The CKM parameters  in Eq.(\ref{eq1}) are  the flavor-changing 
charged current 
couplings of the weak gauge boson $W^{\pm}$ with the quarks as given by 
the Cabibbo-Kobayashi-Maskawa (CKM) unitary quark mixing matrix 
relates the the weak interaction eigenstate $d',s',b'$  of the charge 
$Q=-1/3$ quarks to their mass  eigenstate $d,s,b$ \cite{PDG}:

\be
\kern 1.5cm\pmatrix{d' \cr s' \cr b'}= \pmatrix{V_{ud}& V_{us} & V_{ub} \cr
V_{cd} & V_{cs} & V_{cb} \cr 
V_{td} & V_{ts} & V_{tb}} \pmatrix{d \cr s \cr
b}
\label{Vckm} 
\ee
The unitarity relation 
\be
\kern 1.5cm V_{ud}V_{ub}^{*} + V_{cd}V_{cb}^{*} + V_{td}V_{tb}^{*} =0 
\label{db} 
\ee
can be written as 
\be
\kern 1.5cm R_{b}e^{i\gamma} +  R_{t}e^{-i\beta} = 1
\label{tdb} 
\ee
and is  represented by the $(db)$ unitarity triangle shown in Fig.1. 
$V_{ub} = |V_{ub}|e^{-i\gamma} $, $V_{td} = |V_{td}|e^{-i\beta}$,
and $R_{b} = |V_{ud}V_{ub}^{*}|/|V_{cd}V_{cb}^{*}| $ and 
$R_{t} = |V_{td}V_{tb}^{*}|/|V_{cd}V_{cb}^{*}| $ \cite{Buras}
\begin{figure}[hbp]
\centering
\leavevmode
\epsfxsize=6cm
\epsffile{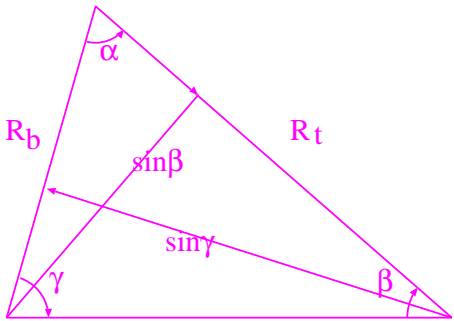}
\caption{The (db) Unitarity Triangle}
\label{fig:db}
\end{figure}

As can be seen from Fig.1, 
\be
\kern 1.5cm R_{b}\sin{\alpha}= \sin{\beta} ,  \quad   R_{t}\sin{\alpha}= \sin{\gamma}
\label{r1} 
\ee
Hence the following lower limits for $R_{b}$ and $R_{t}$ :
\be
\kern 1.5cm R_{b}\geq \sin{\beta},  \quad R_{t}\geq \sin{\gamma}
\label{ll}
\ee
This gives:
\bea
\kern 1.5cm |V_{ud}V_{ub}^{*}| &\geq |V_{cd}V_{cb}^{*}| \sin{\beta} 
\nonumber \\
\kern 1.5cm |V_{td}V_{tb}^{*}| &\geq |V_{cd}V_{cb}^{*}| \sin{\gamma}
\label{lb}
\eea
The current  measured value of $\sin 2\beta$  gives\cite{Nir}
\be
\kern 1.5cm  \sin 2\beta = 0.734\pm 0.06 
\label{bb}
\ee
from which we obtain two solutions for $\sin \beta$:
\be
\kern 1.5cm \sin \beta = \cases { 0.40\pm 0.04 \cr 0.91\pm 0.02 \cr}
\label{beta}
\ee
On the other hand, $|V_{ub}| = (4.11\pm 0.25\pm 0.78)\times 10^{-3} $ 
from LEP and CLEO inclusive data which is above the value 
$(3.25\pm 0.32\pm 0.64)\times 10^{-3} $ from the  CLEO exclusive 
data\cite{PDG}. With the  average value
$|V_{ub}| =  (3.60\pm 0.70)\times 10^{-3} $ from the inclusive and 
exclusive measurements\cite{PDG}, we find 
\be
\kern 1.5cm R_{b} = (0.379\pm 0.07)
\label{Rb}
\ee
which is consistent with the lower limit of $\sin \beta = 0.40\pm 0.04$.
This also excludes the solution $\sin \beta = 0.91$. Conversely, 
from  the measured value for $|V_{cb}| $, a lower limit for $|V_{ub}| $ 
from $\sin\beta$ is then 
\be
\kern 1.5cm |V_{ub}| =  (3.8\pm 0.6)\times 10^{-3}
\label{lVub}
\ee
Note that the inclusive data  is
consistent with the lower bound for $|V_{ub}| $ from $\sin 2\beta$ measurement
while the exclusive data is slighly below the lower bound. Thus to
within experimental errors, the average value for $|V_{ub}| $ suggests 
that $R_{b}\approx \sin\beta$. This implies
\be
\kern 1.5cm \sin\alpha \approx 1, \quad \alpha=\pi/2 ,\quad R_{t} \approx \cos\beta
\ee
\label{rbt}
This is supported by a recent analysis\cite{Buras} which gives:
$\sin 2\alpha =0.05\pm 0.31$ , 
$ R_{b}= 0.404\pm 0.023$ , $R_{t}= 0.927\pm 0.061$
which indeed shows that $\alpha \approx \pi/2$ and $R_{b}\approx \sin\beta$.
If $\alpha $ is known, Eq.(\ref{r1}) can be used to extract $R_{b} $  
and $R_{t} $. We find
\bea
\kern 1.0cm &&R_{b}= \sin\beta\sqrt(1 + (\cos\alpha/\sin\alpha)^{2}),
\nonumber \\
\kern 1.0cm &&R_{t}= \sin\gamma\sqrt(1 + (\cos\alpha/\sin\alpha)^{2})
\label{rbt0}
\eea

This provides a good determination of $|V_{ub}|$ and $|V_{td}|$ in terms of
$\sin\beta $ and $\sin\gamma $, respectively. since $(\cos\alpha)^{2} $
is of the order a few percent. With $\alpha = 90^{\circ}$, 
$\sin\gamma = \cos\beta = 0.91\pm 0.02$, $R_{t}=0.91\pm 0.02$. Thus 
$R_{t}$ is also rather well known. for example, the current determination
of $R_{t}$ from $\Delta M_{d}$ \cite{PDG} gives
$|V_{td}V_{tb}^{*}|= 0.0079\pm 0.0015 , R_{t} = 0.86\pm 0.25 $
which is consistent with the above value for $R_{t}$ from $\sin 2\beta$
measurements. Since $\sin\gamma$
is close to 1, we expect large direct CP violation in $B$ decays. 
We now discuss the CP asymmetry in $B$ decays to 3 pseudo-scalar mesons.

\section{ $B^- \to \pi^+ \pi^- K^-$ and  $B^- \to K^+ K^- K^-$  decays }
Consider now the $B^- \to \pi^+ \pi^- K^-$ and  $B^- \to K^+ K^-
K^-$ decays. The non resonant amplitude is mainly given by the
$O_{1}$, $O_{4}$, $O_{6}$ matrix elements and is dominated by the penguin
operators as in $B\to K\pi$ decays , because of the large CKM factors
relative to the tree amplitude (Fig.2). With
\bea
&&O_{1}= (\bar{s}u)_{L}(\bar{u}b)_{L},  \quad \ O_{4}= \sum_{q}(\bar{s}q)_{L}
(\bar{q}b)_{L}\nonumber \\
&&O_{6}= -2\sum_{q}(\bar{s}_{L}q_{R})(\bar{q}_{R}b_{L})
\label{Oi}
\eea
\begin{figure}[hbp]
\centering
\leavevmode
\epsfxsize=8cm
\epsffile{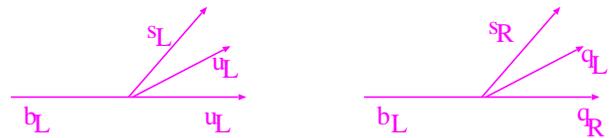}
\caption{Graphs representing the nonleptonic interactions }
\label{fig:b2s}
\end{figure}

In the factorization model, the matrix elements of $O_{1}$, $O_{4}$, $O_{6}$
can be computed in terms of the $B \to \pi\pi l\nu$ form factors given as:
\begin{eqnarray}
&&< \pi^-(p_1) \pi^+(p_2) | ({\bar u} b)_{V -A}
| B^- (p_B)> \mu\nonumber\\
&& = ir(p_B-p_2-p_1)_\mu +iw_+(p_2+p_1)_\mu \nonumber\\
&&+iw_-(p_2-p_1)_\mu - 2h\,\epsilon_{\mu\alpha\beta\gamma}p_B^\alpha p_2^\beta
p_1^\gamma\;.
\end{eqnarray}
\begin{figure}[hbp]
\centering
\leavevmode
\epsfxsize=8cm
\epsffile{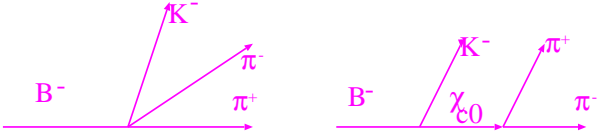}
\caption{Non resonant and resonant $B\to K\pi\pi$ decay }
\label{fig:btokchi}
\end{figure}
The $B^{*}$ pole terms in the form factors $w_{+}$ and $w_{-}$ are obtained
using the Hybrid model (HQET with the 
full propagators for the heavy meson pole terms)\cite{Fajfer2,Bajc2}
\begin{eqnarray}
\kern -0.5cm&&w_+^{nr}(s,t) =  - \frac{g}{f_1 f_2}
\frac{f_{B*} m_{B*}^{3/2} m_B^{1/2}}{(t - m_{B*}^2} [ 1 -   
\frac{(m_B^2 -m_1^2 -t)}{ 2 m_{B*}^2} ]\nonumber\\
\kern -0.5cm && +  \frac{f_B}{ 2 f_1f_2}  -
\frac{{\sqrt m_B} \alpha_2}{ f_1 f_2}
\frac{1}{2 m_B^2}(2 t + s - m_B^2 -m_3^2 -2 m_1^2),\nonumber
\label{w+1-h}
\end{eqnarray}
\begin{eqnarray}
&& w_-^{nr}(s,t)  =   \frac{g}{f_1f_2}
\frac{f_{B*} m_{B*}^{3/2} m_B^{1/2}}{ t  - m_{B*}^2}
[ 1 + \frac{(m_B^2 -m_1^2 -t)}{2 m_{B*}^2}]\nonumber\\
&&  + \frac{{\sqrt m_B} \alpha_1}{f_1f_2}. \nonumber
\label{w-1h}
\end{eqnarray}
We thus have in the factorization model~:
\bea
&&<O_{1}>_{\rm nr} = 
< K^-(p_3) \pi^+(p_1) \pi^-(p_2)| O_1| B(p_B)>_{nr} \nonumber \\
&&= -[f_3 m_3^2 r^{nr} + \frac{1}{2} f_3(m_B^2 - m_3^2 -s)w_+^{nr} \nonumber \\
&&+ \frac{1}{2}f_3 (s + 2 t - m_B^2 - 2 m_1^2 - m_3^2)w_-^{nr}]
\label{o1mv}
\eea
$<O_{4}>_{\rm nr} $ is almost the same as
$<O_{1}>_{\rm nr} $. $<O_{6}>_{\rm nr} $  is given by
\begin{eqnarray}
&&<O_{6}>_{\rm nr} =< K^-(p_3) \pi^+(p_1) \pi^-(p_2)| O_6| B(p_B)>_{nr} \nonumber \\
&&= -(\frac{ {\mathcal B}}{m_B})[ 2 \frac{f_1 f_2}{f_3} m_3^2 r^{nr}
+ \frac{f_1 f_2}{f_3} (m_B^2 + m_3^2 -s)w_+^{nr} \nonumber\\
&&+ \frac{f_1 f_2}{f_3} (s + 2 t - m_B^2 - 2 m_1^2- m_3^2 )w_-^{nr}].
\label{o6mv}
\end{eqnarray}
Thus $<O_{6}>_{\rm nr}= ({2\mathcal B}/m_{B})<O_{1}>_{\rm nr}$ to a good approximation.
The non resonant $B^- \to K^- \pi^+ \pi^-$ amplitude is then 
\bea
&&\kern -0.5cm    {\mathcal M}_{nr} = \frac{G}{{\sqrt 2}}[ V_{ub} V_{us}^* a_1
<O_{1}>_{\rm nr} \nonumber\\
&& - \ V_{tb} V_{ts}^* ( a_4<O_{4}>_{\rm nr}
   + \ a_6<O_{6}>_{\rm nr} )].
\label{amp}
\eea
   \begin{figure}[hbp]
   \centering
\leavevmode
   \epsfysize=5cm
   \epsfbox{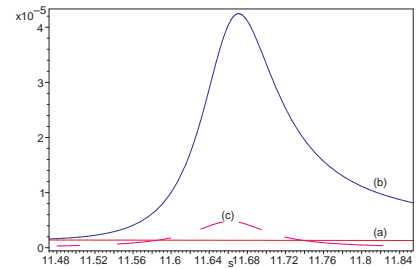}
 \vspace{-1cm}
   \caption{Differential branching ratios for
   $B^{-}\to \pi^{+}\pi^{-}K^{-}$ vs. $s$.
   The curves  (a), (b), (c) are
   $d({\rm B})(\rm NR)/ds$, $d({\rm B})/ds + d(\bar{{\rm B}})/ds$,
     $d({\rm B})/ds - d(\bar{{\rm B}})/ds$
   against the two-pion invariant mass squared s, for the non resonant,
   CP symmetric and CP antisymmetric differential branching ratios
     respectively.}
   \label{Fig.1}
\end{figure}

In the high di-pion invariant mass region, for example at the $\chi_{c0}$
mass, the  two pions in an  $S$-wave $0^{++}$ can also be produced
from  the decay $B^{-} \to \chi_{c0}K^{-}$ followed by the decay
$\chi_{c0} \to \pi^{+}\pi^{-}$ . This will interfere with the non 
resonant amplitude and produce a CP asymmetry at the $\chi_{c0}$
mass. The large  $B^{-} \to \chi_{c0}K^{-}$ branching ratio measured
by Belle and Babar provides a possibility to look for CP violation
in $B \to K\pi\pi$ decays. This large BR also indicates that 
nonfactorizable terms are present in $B$ decays\cite{Colangelo}. The
resonant amplitude is then
\begin{eqnarray}
\kern -0.7cm &&{\mathcal M}_{r}(B^- \to  \chi_{c0} K^- \to \pi^+ \pi^- K^-)  = \nonumber\\
\kern -0.7cm &&{\mathcal M}(B^{-} \to \chi_{c0} K^- ) \frac{1}{ s - m_{\chi_{c0}}^2 +  i
\Gamma_{\chi_{c0}} m_{\chi_{c0}}}
{\mathcal M}( \chi_{c0} \to \pi^+ \pi^- ).
\label{ares}
\end{eqnarray}
Using the Belle measured $B^- \to  \chi_{c0} K^- $ branching 
ratio\cite{Belle1}, we
computed the differential decay rates and CP asymmetries for
the two-pion and two-kaon system in the $\chi_{c0} $ mass region shown
in Figs.1 and Figs.2  (taken from the published work\cite{Fajfer1}).
For the integrated branching ratios we obtain:
\be
{\rm BR}(B^-\to K^- \pi^+ \pi^-)_{nr} = T + P + I_1 \cos \gamma
   + I_2 \sin \gamma. \nonumber
\label{kpp-b1}
\ee
with
$T = 7.0\times 10^{-6}$, $P= 7.5 \times 10^{-5}$,
$I_1 = -4.3 \times 10^{-5}$, $I_2 = -1.5 \times 10^{-5}$, For 
$\sin\gamma=0.9$, $\cos\gamma=0.4$, $g=0.23$, we obtain a branching ratio
 ${\rm BR}(B^-\to K^- \pi^+ \pi^-)_{nr}=5.1 \times 10^{-5} $ 
for the non resonant $B^-\to K^- \pi^+ \pi^- $ decay. 
The effective $DD^{*}\pi$ coupling $g$ used here is smaller than the soft-pion
value of $0.57$ because of off-shell effects. $g$ and other parameters
$\alpha_{1}$ and $\alpha_{2}$ are chosen to fit the $B\to \pi$
and $B\to \rho$ form factors. The above BR is consistent with the 
experimental upper limits for the non resonant BR. Similarly,
\be
{\rm BR}(B^-\to K^- K^+ K^-)_{nr} = T + P + I_1 \cos \gamma
   + I_2 \sin \gamma. \nonumber
\label{kkk-b1}
\ee
and
 $T = 3.4\times 10^{-6}$, $P= 3.7\times 10^{-5}$, 
$I_1 = -2.1 \times 10^{-5}$ ,$I_2 = -7.4 \times 10^{-6}$. This gives
${\rm BR}(B^-\to K^- K^+ K^-)_{nr} = 2.53\times 10^{-5} $ for the 
non resonant $B^-\to K^- K^+ K^- $ decay.
The CP asymmetry of the total decay rate is given by 
\be
\kern 2.0cmA = \frac{ \sin \gamma N_1}{N_2 + \cos \gamma N_3}, 
\label{asint1}
\ee
where
$N_1= -3.0 \times 10^{-5}$, $N_2= 16.4 \times 10^{-5}$ , 
$N_3= -8.6\times 10^{-5}$ for $B^- \to  \pi^+ \pi^- K^-$ 
and $N_1= -1.5 \times 10^{-5}$, $N_2= 8.2 \times 10^{-5}$,
$N_3= -4.2\times 10^{-5}$ for $B^- \to  K^+ K^- K^-$~.
and the total integrated asymmetry is then
\bea
\kern 1.0cm &&A(B^- \to  \pi^+ \pi^- K^-) = -0.20\nonumber \\
\kern 1.0cm &&A(B^- \to  K^+ K^- K^-)=-0.20 
\label{acp}
\eea
Since $<O_{4}>_{\rm nr}=<O_{1}>_{\rm nr}$ and
$<O_{6}>_{\rm nr}$ is  proportional to $<O_{1}>_{\rm nr}$
the above results for the non resonant CP 
asymmetry are
essentially independent of the form factors $w_{+}$ and $w_{-}$. 
For the  partial width CP asymmetry integrated over the 
the $\chi_{c0}$ mass region, we find  
\bea
\kern -0.2cm && A_p(B^\pm\to K^\pm \pi^+ \pi^-) = 7.9 \sin \gamma/(73 - 1.2 \cos \gamma)\nonumber \\
\kern -0.2cm && A_p(B^\pm\to K^\pm K^+ K-) =  7.2 \sin \gamma/(41 - 5.6 \cos \gamma ) 
\eea
   \begin{figure}[hbp]
   \centering
\leavevmode
   \epsfysize=5cm
   \epsfbox{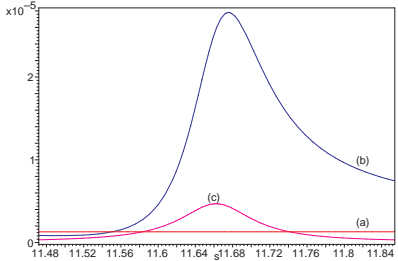}
 \vspace{-1cm}
   \caption{Differential branching ratios for
   $B^{-}\to K^{+}K^{-}K^{-}$ vs. $s$.
   The curves  (a), (b), (c) are
   $d({\rm B})(\rm NR)/ds$, $d({\rm B})/ds + d(\bar{{\rm B}})/ds$,
     $d({\rm B})/ds - d(\bar{{\rm B}})/ds$
   against the two-kaon invariant mass squared s, for the non resonant,
   CP symmetric and CP antisymmetric differential branching ratios
     respectively.}
   \label{Fig.2}
   \end{figure}

\bigskip
I would like to thank the Organisers of the CKM03 Workshop for
the warm hospitality at Durham  and the \break EURIDICE Collaboration
for a grant.

\end{document}